\documentclass[epj]{webofc}
\usepackage[utf8]{inputenc}
\usepackage[varg]{txfonts}   % Web of Conferences font
\usepackage{booktabs}
\usepackage{xcolor}
\definecolor{darkred}{rgb}{0.4,0.0,0.0}
\definecolor{darkgreen}{rgb}{0.0,0.4,0.0}
\definecolor{darkblue}{rgb}{0.0,0.0,0.4}
\usepackage[bookmarks,linktocpage,colorlinks,
    linkcolor = darkred,
    urlcolor  = darkblue,
    citecolor = darkgreen]{hyperref}
\usepackage{subfigure}
\wocname{EPJ Web of Conferences}
\woctitle{Lattice2017}
\newcommand{\ignore}[1]{}

\newcommand{\esp}[1]{\langle #1 \rangle}
\newcommand{\C}{\mathbb{C}}
\newcommand{\R}{\mathbb{R}}
\newcommand{\Z}{\mathbb{Z}}
\renewcommand{\Im}{\mathrm{Im}}
\renewcommand{\Re}{\mathrm{Re}}
\newcommand{\alert}[1]{{\em #1}}
\begin{document}
%%%%%%%%%%%%%%%%%%%%%%%%%%%%%%%%%%%%%%%%%%%%%%%%%%%%%%%%%%%%%%%%%%%%%%%%%%%%%
%
\selectlanguage{english}
%----------------------------------------------------------------------------
\title{%
  Representation of complex probabilities and complex Gibbs sampling
}
%----------------------------------------------------------------------------
\author{%
\firstname{Lorenzo Luis}
\lastname{Salcedo}
\inst{1,2}
\fnsep\thanks{Acknowledges financial support by
Spanish Ministerio de Econom\'{\i}a y Competitividad (Grant
No. FIS2014-59386-P), and Agencia de Innovaci\'on
y Desarrollo de Andaluc\'{\i}a (Grant No. FQM225),
and Centro de Servicios de
Informática y Redes de Comunicaciones (CSIRC),
Universidad de Granada, for providing computing time,
\email{salcedo@ugr.es}}
}
%----------------------------------------------------------------------------
\institute{%
Departamento de F\'{\i}sica At\'omica Molecular y Nuclear,
Universidad de Granada, E-18071, Spain
\and
Instituto Carlos I de F\'{\i}sica Te\'orica y Computacional,
Universidad de Granada,  E-18071, Spain
}
%----------------------------------------------------------------------------
\abstract{ Complex weights appear in Physics which are beyond a
  straightforward importance sampling treatment, as required in Monte Carlo
  calculations. This is the well-known sign problem. The complex Langevin
  approach amounts to effectively construct a posi\-tive distribution on the
  complexified manifold reproducing the expectation values of the observables
  through their analytical extension. Here we discuss the direct construction
  of such positive distributions paying attention to their localization on the
  complexified manifold. Explicit localized repre\-sentations are obtained for
  complex probabilities defined on Abelian and non Abelian groups. The
  viability and performance of a complex version of the heat bath method,
  based on such representations, is analyzed. }
%----------------------------------------------------------------------------
\maketitle

%\tableofcontents

%----------------------------------------------------------------------------
\section{Motivation and existing approaches}

Many physical problems reduce to computing weighted averages
\begin{equation}
\esp{A}_P = \frac{\int d^nx \, P(x)\, A(x)}{\int d^nx \, P(x)}
\qquad
x\in \R^n
\qquad
P(x) = e^{-S(x)}
.
\end{equation}
For a large number of entangled degrees of freedom the \alert{Monte Carlo}
method is often the only viable approach:
\begin{equation}
x \sim P \qquad \text{(\alert{Importance sampling} with respect to $P(x)$)}
.
\end{equation}
However \alert{complex probabilities} do appear in certain cases. An
outstanding example is lattice QCD at finite density \cite{Hasenfratz:1983ba}.
Straightforward importance sampling is meaningless when $P(x)$ is not a
positive distribution and this is the well known \alert{sign problem}
\cite{Troyer:2004ge}.

Several proposals have been put forward to deal with complex weights within
Monte Carlo, among other: Reweighting, complex Langevin equation (CLE)
\cite{Parisi:1984cs,Klauder:1983nn,Aarts:2009uq,Nagata:2015uga,Salcedo:2016kyy,SWtc1},
Lefschetz thimbles and variants \cite{Cristoforetti:2012su,Alexandru:2015sua},
reweighting the Complex Langevin equation \cite{Bloch:2011jx}, or particular
treatments for special problems, e.g., worm algorithms
\cite{Prokofev:2001ddj}, or other \cite{Wosiek:2015bqg}.

%\note{CITAR AQUI mi ejemplo de CLE failing y overlap en RW}

%\subsection{The goal}

The sign problem is hard \cite{Troyer:2004ge}. Nothing as robust and reliable
as the Metropolis algorithm is avai\-lable in the complex case and each of the
abovementioned methods has well-known limitations or is too recent to asses
its performance. It is fair to say that, at present, no existing method can be
regarded as the complete solution. In this view new alternative approaches are
worth pursuing.

Our goal is to explore a complex version of the \alert{heat bath} method. The
idea is to do the updates by means of a \alert{representation} of the
conditional \alert{complex} probability. This requires studying
representations of complex probabilities by themselves (i.e., quite
independently of the CLE construction).

\section{Direct positive representations}

\subsection{Representations by themselves}

%\subsubsection{Direct representations. I}
\subsubsection{Direct representations}

The CLE is based on representing $\,P(x)= e^{-S(x)}\,$ by an ordinary
(real and positive) probability distribution $\rho(z)$ defined on the
\alert{complexified manifold}, so that ideally $\esp{A}_P = \esp{A}_\rho $
where $A(z)$ is the \alert{analytical extension} of $A(x)$.  In the CLE the
distribution $\rho(z)$ is not obtained nor needed explicitly.  Abstracting the
idea, one can define \cite{Salcedo:1996sa} a \alert{representation} $\rho(z)$
of a complex distribution $P(x)$ on $\R^n$, as a distribution on $\C^n$ such
that
\begin{equation}
\int d^{2n}z\, \rho(z)\, A(z) = \int d^{n}x\, P(x)\, A(x) 
\qquad
\text{for generic $A(x)$}
.
\end{equation}
Equivalently,
\begin{equation}
P(x) = (K\rho)(x) \equiv \int d^ny \, \rho(x - i y , y)
\quad\text{(\alert{analytical projection})}
.
\end{equation}
Of interest for Monte Carlo are the \alert{positive representations}, that is,
those with $\rho(z)\ge 0$.

As it turns out, positive representations exists very generally and
constructions are available for many complex probabilities, including Gaussian
times polynomial of any degree and in any number of dimensions (a dense set)
\cite{Salcedo:1996sa}, periodic distributions in any number of dimensions,
complex distributions on compact Lie groups and coset spaces
\cite{Salcedo:2007ji}. In such representations $P(x)$ needs not be an
analytical function, any (normalizable) complex distribution has positive
representations sharing the same symmetries enjoyed by the complex probability
itself. Moreover, the expectation values of the subset of the holomorphic
function do not fully fix $\rho(z)$: Representations of a given $P(x)$ are by
no means unique. E.g., if $\rho$ is a positive representation, so is
~$\exp(\,t\nabla^2) \rho ~~ (t>0)$ \cite{Salcedo:1996sa}.

\subsubsection{A concrete construction}

Given a complex probability $P(x)$ on $\R^n$ let us choose $P_0(x)>0$ and
$H(x)$ such that
\begin{equation}
\int d^nx\, P(x) = \int d^nx\, P_0(x) 
,\qquad
P(x) = P_0(x) + \nabla(\, P_0(x) \, H(x)\,)
.
\end{equation}
Then it is easy to verify \cite{Salcedo:2007ji} that $\esp{A(z)}=\esp{A}_P$
with $z = x^\prime + z^\prime H(x^\prime)$, where $x^\prime\sim P_0$ and
$z^\prime \sim q$, and $q(z)$ is any positive representation of
$\,\delta(x)+\delta^\prime(x)$, ~such as $\displaystyle q(z) =
\frac{1}{8\pi}\left| 1-\frac{z}{2}\right|^2 e^{-{|z|^2/4}}$.

Nevertheless, the sign problem is not solved in practice since the available
constructions are not viable for large systems: $(i)$ the normalization of
$P(x)$ is not usually known, $(ii)$ the vector field $H(x)$ is not easy to
obtain for large $n$, and $(iii)$ the dispersion of $z$ increases
exponentially with $n$, spoiling the Monte Carlo estimates.

\subsubsection{Conditions on the support of positive representations}

This is an important point in the representation approach: representations
with smaller widths yield smaller variances and poor representations may not
be better than reweighting.  As a rule, the more complex $P(x)$, the wider
must be $\rho(z)>0$.  For given $P(x)$, the \alert{width} (size of the support
in the imaginary direction) of any positive $\rho(z)$ is bounded from
below. For instance, the relation
\begin{equation}
|\esp{A(x)}_P| = | \esp{A(z)}_\rho | \le
\max_{z \, \in \, {\rm supp} \,\rho} \{ | A(z) | \}
\quad
\text{for arbitrary $A$}
\end{equation}
constrains how localized $\rho$ can be. In particular $A=e^{-ikx}$ provides
quite explicit bounds \cite{Salcedo:2015jxd}.

For local actions good quality positive representations (i.e., with bounded
width as $n\,$ increases) should exist, eventually outperforming reweighting:
CLE being an example, when it works.

\subsection{Two-branch representations}

%\subsubsection{Two-branch representations: one-dimensional case. I}
\subsubsection{Two-branch representations: one-dimensional case}

\begin{figure}[ht]
%\sidecaption
\centering
\includegraphics[height=40mm]{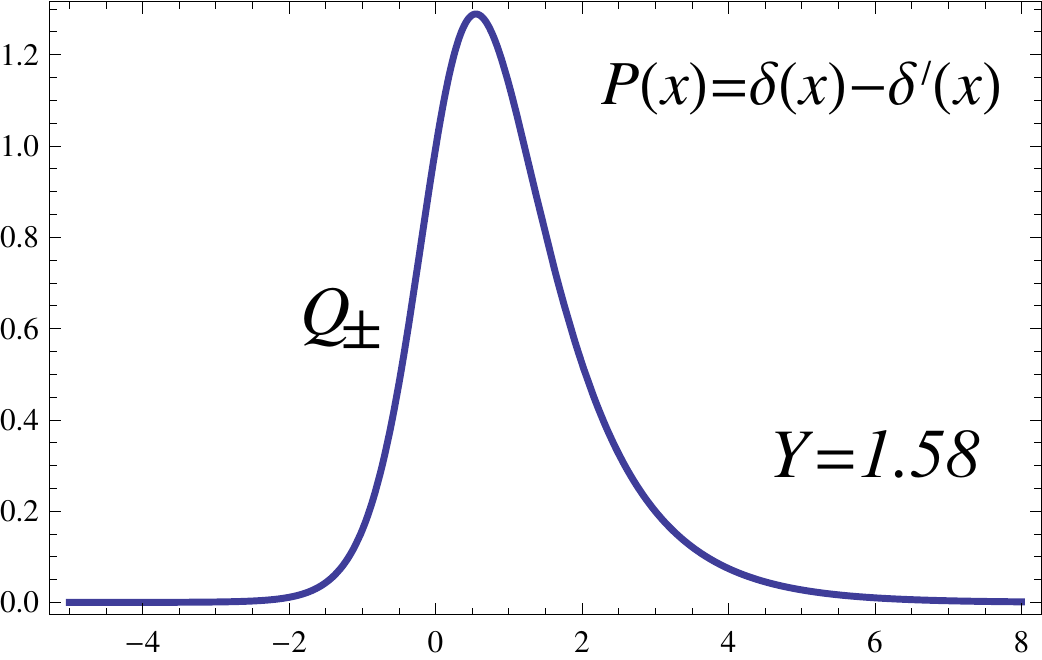}
\includegraphics[height=40mm]{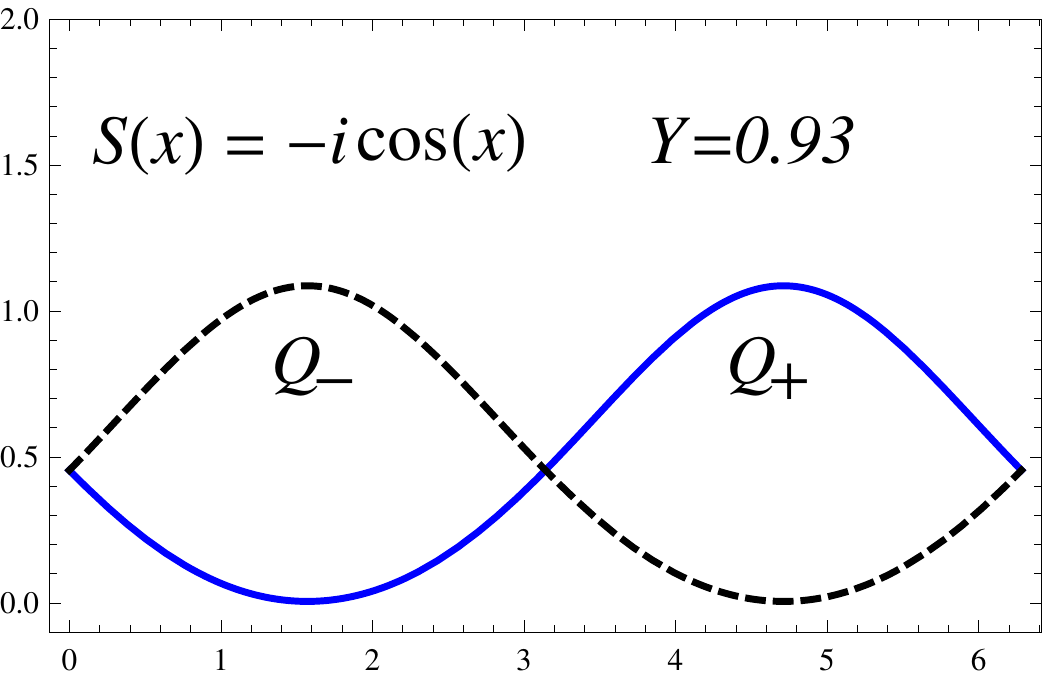} 
  \caption{Two examples of one-dimensional two-branch representations.}
  \label{fig:twob-2}
\end{figure}

Attending to the issue of the width, let us consider representations with
support along two horizontal lines, $\R\pm i Y$ ~(\alert{two-branch
  representations} \cite{Salcedo:2015jxd}):
\begin{equation}
\rho(z) = Q_+(x) \, \delta(y-Y) +
Q_-(x) \, \delta(y+Y) \qquad Q_\pm(x) \ge 0 
.
\end{equation}
It is easy to check that such $\rho(z)$ will represent $P(x)$ provided
\begin{equation}
P(x) = Q_+(x-iY) + Q_-(x+iY)
,
\end{equation}
due to 
\begin{equation}
\esp{A(x)}_P
= \sum_{\sigma=\pm} N_\sigma \esp{A(x+i\sigma Y)}_{Q_\sigma}
= \esp{A(z)}_\rho
,\quad
N_\sigma \equiv \int dx \, Q_\sigma(x)
.
\end{equation}
For any $Y$, \alert{real} functions $\,Q_\pm(x)$ exist and they are
essentially unique. In terms of Fourier modes:
\begin{equation}
\tilde Q_\pm(k)
=
\pm \frac{ e^{\pm k Y} \tilde{P}(k) -  e^{\mp k Y} \tilde{P}(-k)^* }{2 \sinh(2kY)}
\quad (k\not=0)
\end{equation}
with $P(x) = \sum_k e^{ikx} \tilde{P}(k)$.  For $Y$ above some critical value,
the $Q_\pm(x)$ become \alert{positive}: there are more $e^Y$ in denominator
and eventually the positive constant mode dominates. The critical $Y$ is the
optimal choice since it has the smallest width. Two examples are displayed in
figure~\ref{fig:twob-2}.

%charlas/lattice35/mth/5.nb

\subsubsection{Two-branch representations: higher-dimensional case}

For complex probabilities defined on $\R^n$ ($n>1$) one can try a
\alert{strict} two-branch scheme \cite{Salcedo:2015jxd} with
\begin{equation}
P(\vec{x}) = \sum_{\sigma = \pm} Q_\sigma (\vec{x} -i \sigma \vec{Y})
,
\qquad
Q_\sigma (\vec{x}) \ge 0 
,
\end{equation}
\begin{equation}
Q_\sigma (\vec{x}) = N_\sigma + 2 \Re \sum_{\vec{k}\not=0}
\frac{ e^{i \vec{k} \cdot (\vec{x} -i\sigma \vec{Y})} \tilde{P}(k) }
{2 \sinh(2\sigma \vec{k} \!\cdot\! \vec{Y})}
 \equiv
N_\sigma + \Re \left( C(\vec{x}; \sigma \vec{Y}) * P(\vec{x}) \right)
.
\end{equation}
A similar construction has been proposed in \cite{Seiler:2017vwj,SWtc}.  Once
again, this is a \alert{formal} solution but not a \alert{practical} one. In
addition there is a technical problem: there are no good choices for
$\vec{Y}$. The natural choice $\vec{Y}= (Y,\ldots,Y)$ meets $\vec{k} \!\cdot\!
\vec{Y} = 0$ for some Fourier modes because these components are not moved
into the complex manifold. Asymmetric $\vec{Y}$ are unnatural and also tend to
require some $Y_i$ larger than necessary.  An elegant solution is to use $2^n$
branches: For each Fourier mode $\vec{k}$ and variable $x_i$, either $+Y$ of
$-Y$ is selected so that $k_i Y_i \ge 0$ for $i=1,\ldots,n$:
\begin{equation}
P(x) = \sum_{\vec{\sigma}} Q_{\vec{\sigma}} (\vec{x} -i \vec{\sigma} Y)
\qquad
\vec{\sigma}  =(\pm,\ldots,\pm)
\end{equation}
Effectively one is adding a bit for each degree of freedom (say, each site),
from $\vec{x} \in \R^n$ to $ (\vec{x},\vec{\sigma}) \in  (\R
\times \Z_2)^n \,$.  Having more branches is compensated by i)~simplicity, ii)
restoration of symmetry, iii) smaller values of $Y$, and iv) it is mandatory
in the non periodic case.

\begin{figure}[thb]
\sidecaption
  \centering
  \includegraphics[height=35mm]{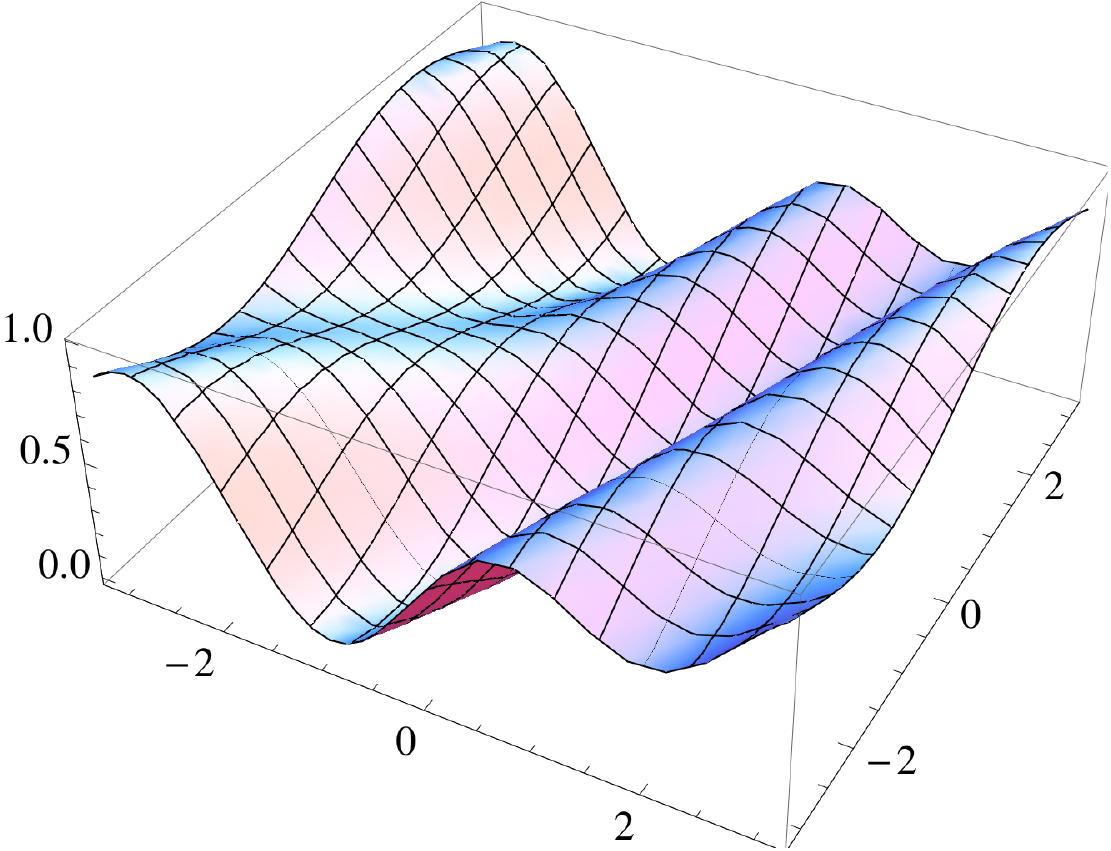}
  \includegraphics[height=35mm]{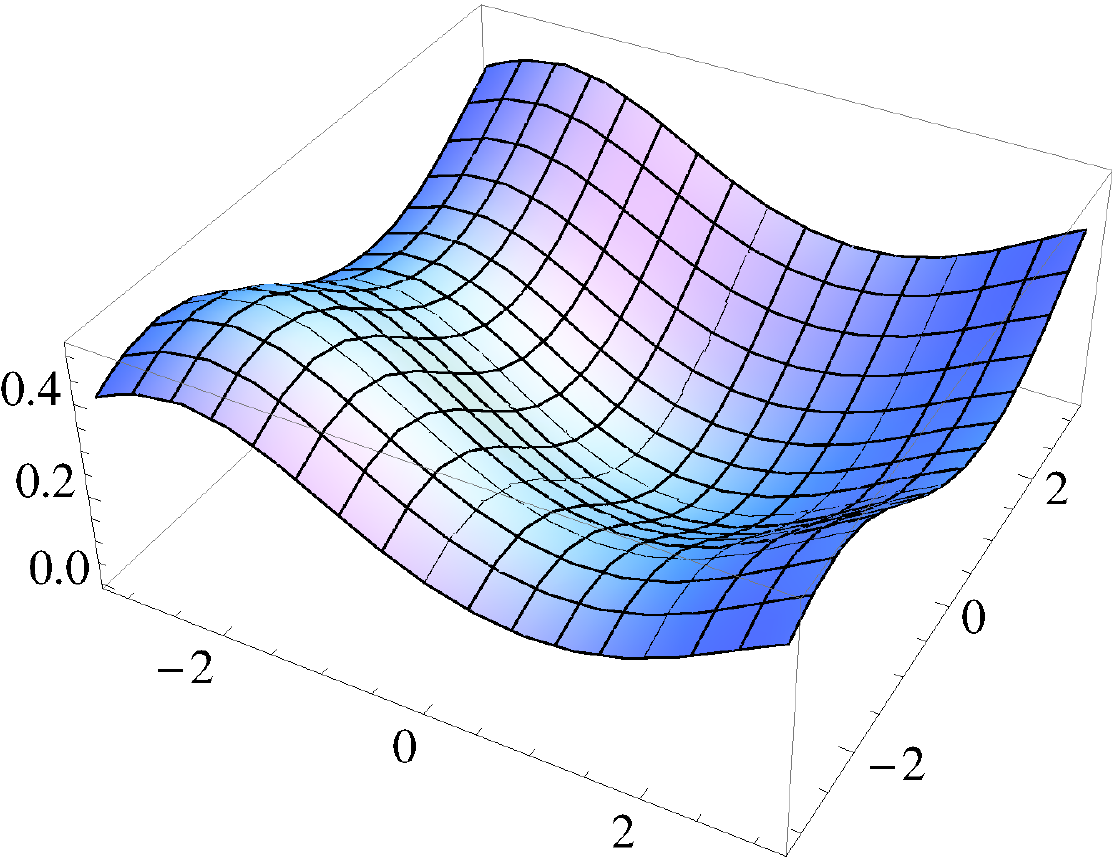}
  \caption{Left: (Representation using 2 branches) $Q_+(\vec{x})$, ~ $\vec{Y}
    = (1.48,4.43)$.  Right: (Representation using $2^n$ branches)
    $Q_{++}(\vec{x})$, ~$\vec{Y} = (2.4,2.4)$.}
  \label{fig:twob-3}
\end{figure}
Figure~\ref{fig:twob-3} displays examples of strict and generalized
two-branch representations in two dimensions for
\begin{equation}
P( x_1 , x_2 ) = (1 + \beta \cos(x_1) ) (1 + \beta \cos(x_2) )
(1 + \beta \cos(x_1 - x_2) )
,
\quad
\beta = 2i
.
\end{equation}

\subsection{Representation of complex probabilities on groups}

%\subsubsection{Representation of complex probabilities on groups. I}

Let $P(g)$ be a complex probability defined on a matrix group $G = \{
e^{-i\vec{a}\cdot\vec{T}}, ~\vec{a} \in \R^m\} $. The observables can be
analytically extended to the \alert{complexified group} $\tilde{G} = \{
e^{-i\vec{a}\cdot\vec{T}}, ~\vec{a} \in \C^m\} $, Haar measures can be adopted
on $G$ and $\tilde{G}$ and representations can be defined as before.

The \alert{two-branch scheme} can be adapted to the present case: Letting
~$G_I \equiv \{ e^{-i\vec{a}\cdot\vec{T}}, ~\vec{a} \in i\R^m\} $,
\begin{equation}
P(g) = Q_+(gh) + Q_-(gh^{-1}),
\quad
g\in G, \quad
h \in G_I
,
\label{eq:18}
\end{equation}
\begin{equation}
\esp{A}_P = N_+ \esp{A(gh^{-1})}_{Q_+} + N_- \esp{A(gh)}_{Q_-}
= \esp{A}_\rho
,
\end{equation}
where $A(gh^{-1})$, $A(gh)\,$ refer to the analytical extension in
$\tilde{G}$. So the Monte Carlo sampling is carried out using the real and
positive weights $Q_\pm(g)$ defined on $G$.

To do the matching in Eq. (\ref{eq:18}), $P$ and $Q_\pm$ are expanded as a
linear combination of \alert{group representations} (generalizing the Fourier
modes of ${\mathrm U}(1)$).  For instance in ${\mathrm U}(N)$, for
\begin{equation}
P(g) = 
\sum_n a^{\, j_1\cdots j_n}_{\, i_1 \cdots i_n} \, g^{\,i_1}{}_{j_1}\cdots g^{\,i_n}{}_{j_n}
\label{eq:17}
\end{equation}
(not the most general case) one finds the solution, choosing $\,h =
\mathrm{diag} ( \omega_1, \ldots, \omega_N )$ ~~($\omega_i > 0$),
\begin{equation}
Q_\pm(g) =
2 \,\Re \sum_n a^{\, j_1\cdots j_n}_{\, i_1 \cdots i_n} \, g^{\,i_1}{}_{j_1}\cdots
 g^{\,i_n}{}_{j_n}
\, \chi( \Omega_{j_1,\ldots,j_n}^{\pm1} )
,
\qquad
\chi(\Omega) \equiv \frac{\Omega}{\Omega^2 - \Omega^{-2}}
,
\label{eq:18a}
\end{equation}
where ~$\Omega_{j_1,\ldots,j_n} \equiv \omega_{j_1} \!\cdots
\omega_{j_n}$. Once again, for sufficiently large $h\,$ the constant (group
singlet) mode dominates and $Q_\pm > 0$. (The constant mode is to be added
  separately both in Eq.~(\ref{eq:17}) and in Eq.~(\ref{eq:18a}).)

\begin{figure}[thb]
\sidecaption
  \centering
  \includegraphics[width=60mm,clip]{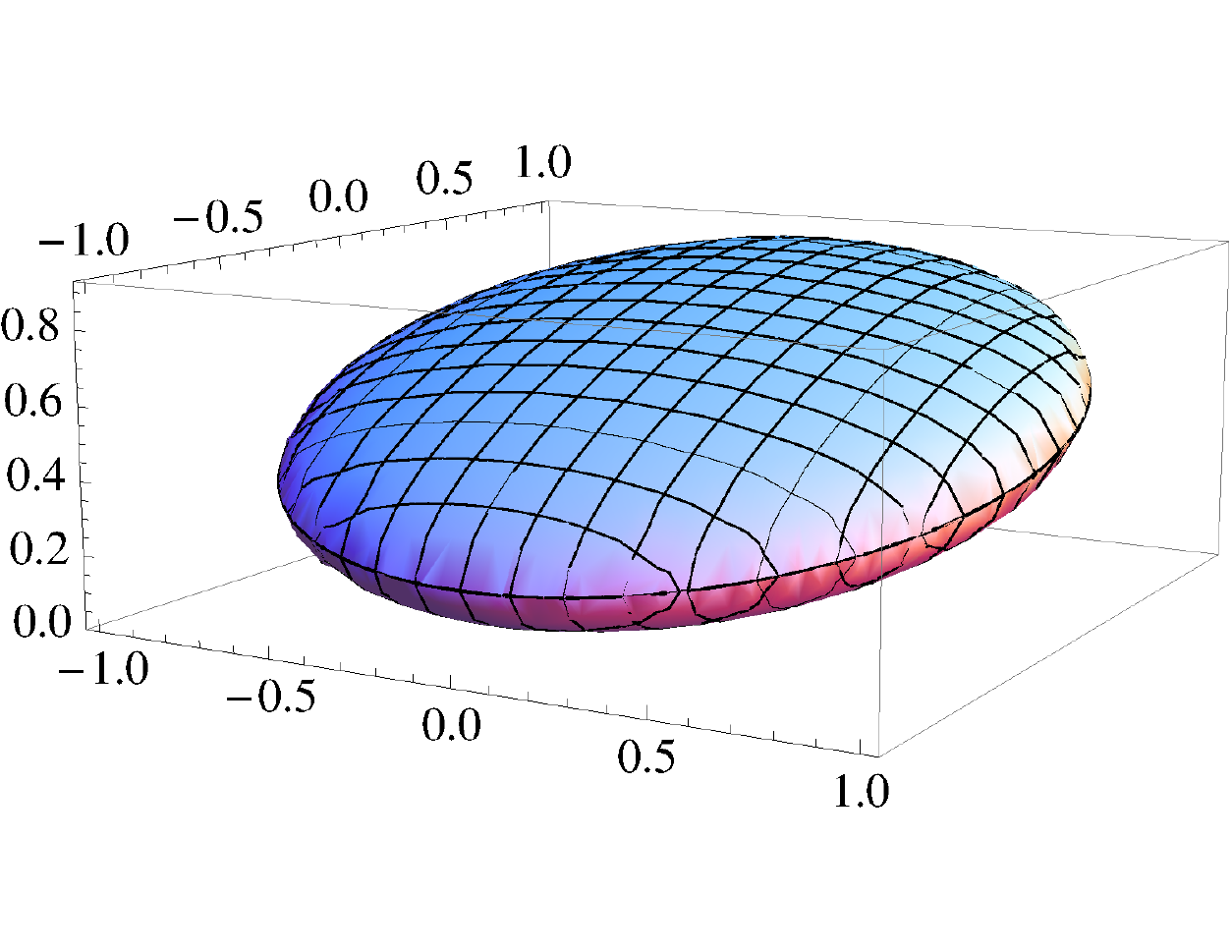}
  \caption{For $P(g) = 1 + \beta \, \mathrm{tr}(\sigma_3 \, g)$ in
    $\mathrm{SU}(2)$, the function $Q_+(g)$ is displayed for $a_2=0$ on the
    plane $(a_1,a_3)$, where $g=a_0-i\vec{a}\cdot\vec{\sigma}$. In the plot
    $\beta = \frac{1}{2}(1+i)$ and $\Omega = e^Y$, $Y=2.5$. The function is
    two-valued corresponding to $a_0=\pm\sqrt{1-\vec{a}\,{}^2}$ ($a_2=0$
      defines a manifold $\mathrm{S}^2$ within $\mathrm{S}^3 \cong
      \mathrm{SU}(2)$ ). }
  \label{fig:su2}
\end{figure}
As an example, a two-branch representation on $\mathrm{SU}(2)$ for ~$P(g) = 1
+ \beta \, \mathrm{tr}(\sigma_3 \, g)$ ~is
\begin{equation}
Q_\pm(g) = \frac{1}{2} + \mathrm{tr}\! \left[\left(
\pm\frac{\beta_R}{\omega-\omega^{-1}} + i\sigma_3
\frac{\beta_I}{\omega+\omega^{-1}} \right) g \right]
,
\qquad
h = \mathrm{diag}(\omega,\omega^{-1})
.
\end{equation}
$Q_+(g)$ is displayed in figure~\ref{fig:su2}, using
$g=a_0-i\vec{a}\cdot\vec{\sigma}$, ~and~ $(a_0,\vec{a}) \in \mathrm{S}^3$.

As it would be expected subtleties arise in the non-Abelian case: regardless
of the choice of $h$, certain components (singlet with respect to $h\,$) will
remain unchanged (would have $\Omega=1$ in Eq.~(\ref{eq:18a})). Equation
(\ref{eq:18}) can only be fulfilled if $P$ is real along those components. For
them a different element $h^{\,\prime}\in G_I$ has to be applied, subject to
the condition
\begin{equation}
\esp{\, \text{singlet}~h \,| \,\text{singlet}~h^{\,\prime} \, } = 0
.
\end{equation}

\section{Complex heat bath approach}

\subsection{The complex Gibbs sampling method}

The construction of direct representations just discussed becomes rapidly
inviable as the dimension of the manifold increases. This suggests a heat bath
approach. As in the standard Gibbs sampling, each variable is updated in turn,
but, since the conditional probability is complex, a
\alert{representation} of it is used instead:
\begin{equation}
(z_1,\ldots,z_i,\ldots,z_n) \to (z_1,\ldots,z_i^\prime,\ldots,z_n), 
\quad
z_i^\prime \sim \rho_{\text{rep}}(z_i^\prime| \{z_{j\not= i}\} ) 
\end{equation}
where $ \rho_{\text{rep}}(z_i^\prime| \{z_{j\not= i}\} ) $ ~is a
representation (with respect to $z_i^\prime$) of the conditional probability
\begin{equation}
P(z_i^\prime | \{ z_{j\not=i} \} ) = 
\frac
{P(z_1,\ldots,z_i^\prime,\ldots,z_n)}
{P(z_1,\ldots,\widehat{z_i},\ldots,z_n)}
.
\end{equation}
By construction only one-variable representations are needed, however, the
method relies on the analytical extension of $P({\vec x})$, just like CLE.
Furthermore convergence to a representation of $P(x)$ is not guaranteed
as standard Markovian chain theorems do not apply. Also, there is a potential
problem from zeroes of the \alert{marginal} probabilities
$P(z_1,\ldots,\widehat{z_i},\ldots,z_n)$.

To see the performance of the method we have studied a concrete
model~\cite{Salcedo:2015jxd}, namely, a complex action in a $d$-dimensional
periodic lattice
\begin{equation}
S[\phi] = \sum_x \! \Big(
\lambda \phi_x^4 + \phi_x^2 + \beta \, \phi_x \sum_{\mu=1}^d \phi_{x+\hat{\mu}}
\Big)
,\quad
\beta \in \C
\quad \lambda \ge 0
,
\label{eq:26}
\end{equation}
with conditional probability
%\begin{equation}
$
P[\phi_x| \{\phi_{x^\prime \not= x}\}]
\, \propto \,
e^{
-\lambda \phi_x^4 - \phi_x^2 - \beta \, \phi_x \sum_{\mu=1}^d
(\phi_{x+\hat{\mu}}
+ \phi_{x - \hat{\mu}})
}
.
$
%\end{equation}
In the \alert{free} case ($\lambda=0$), this conditional probability is a
displaced real Gaussian and the marginal probability has no zeroes. The method
works alright for a bounded region of the $\beta$ plane ($e^{-S}$ is
normalizable for bounded $\Re(\beta)$).

\subsection{Monte Carlo study of an interacting case}

We further apply the complex Gibbs sampling approach for $\lambda=1$ and
imaginary $\beta$. Some checks from $T$-matrix calculations, for $d=1$ only,
show that the method works up to $|\beta| \approx 1$.

For higher dimensions we have used checks from virial (or Schwinger-Dyson)
relations
\begin{equation}
%$\displaystyle
\left\langle \frac{\partial A}{\partial \phi_x} \right\rangle
 = 
\left\langle A \frac{\partial S}{\partial \phi_x} \right\rangle
.
%$
\end{equation}
The choices $A=\phi_x$ and $A=\phi_{x+\hat\mu}\,$, give rise to four
observables $O_{1,2}$ and $I_{1,2}$ such that $\esp{I_1} = \esp{I_2} = 0.$
These observables are computed using Monte Carlo and results are displayed in
table~\ref{tab:1}.  Reweighting and complex Langevin results are also
displayed for comparison.%\footnote{The author thanks
%  to CSIRC (Granada University) for providing computing time in the cluster
%  Alhambra.}

\newcommand{\fm}{{\phantom{$-$}}}
\newcommand{\fz}{{\phantom{0}}}
\newcommand{\fmb}{\hspace{3.1mm}}

%\small\scriptsize

\begin{table}[thb]
\small
\centering
\caption{Some Monte Carlo results for the action in Eq.~(\ref{eq:26}) with
  $\lambda=1$ and various geometries and couplings \cite{Salcedo:2015jxd}. The
  labels RW, CLE and CHB stand for the method used: reweighting, complex
  Langevin equation, and complex heat bath (with a cutoff),
  respectively. CHB$^*$: no cutoff. CHB$^{**}$: Filtered.}
\label{tab:1}
\begin{tabular}{crlllll}\toprule

$\beta$ & $N^d$ %& $L$ & $K$ 
& $ 10^3 \! \times \!\Re\langle O_1 \rangle $
& $ 10^3 \! \times \! \Im\langle O_2 \rangle $
& $ 10^3 \! \times \! \Re\langle I_1 \rangle $
& $ 10^3 \! \times \! \Im\langle I_2 \rangle $
& Method
\\ 
\hline

%{{51, rwmnlgp}}Infinity
0.25i & $3^3$ % &  & 
 & 19.783 \fz (28) %&  $-$ i\, 1.134 \fz (33)
% & \fm 2.984 (12) 
&    56.017 (19)
 & \fm 0.49 \fz (33) %&  $-$ i\, 0.07 \fz \fz (9)
% &   $-$ 0.22 (10) 
&  $-$ 0.05 \fz (8)
 & RW
\\

%{{2031, clnlgp}}Infinity
0.25i & $3^3$ % &  & 
 & 19.740 \fz (21) %&  $-$  1.120 \fz (99)
% & \fm 3.009(17) 
&    55.978 (51)
 & \fm 0.97 \fz (73) %&  $-$  0.09 \fz\fz (8)
% & $-$ 0.10  (30) 
&  \fmb  0.02 \fz (4)
 & CLE
\\

%{{102, nlgp1}, 3, 3, 0, 100, 1000, 1000, 40626, 40626, 0.  0.25 I, 4, 1000, 1., 20., 10, 5.}Infinity
0.25i & $3^3$ % & 20 & 10
 & 19.745 \fz (14) %&  $-$  1.099 \fz (28)
% & \fm 2.996(11) 
&    55.967 (16)
 & \fm 3.39 (115) %&    0.09 \fz (23)
% & $-$0.12 \fz (9) 
&  $-$  0.14 \fz (9)
 & CHB
\color{black}
\\
\hline

%{{55, rwnlgp}, 3, 8, 0, 100, 1000, 1000, 89498, 89498, 0.  0.25 I, 4, 1000, 1., 20., 10}Infinity
0.25i & $8^3$ % &  &  
 & 19.985 (137) %&  $-$  0.153 (179)
% & $-$ 0.074 (47) 
&    55.958 (44)
 & \fm 0.59 \fz (63) %&    1.16 \fz (68)
% & $-$ 0.55 (53) 
&  $-$  0.67 (42)
 & RW
\\

%{{2035, clnlgp}}Infinity
0.25i & $8^3$ % &  & 
 & 19.746 \fz\fz (5) %&  $-$  0.021 \fz(20)
% & \fm 0.004 \fz (4) 
&    55.977  (12)
 & \fm 0.29 \fz (24) %&  $-$  0.02 \fz\fz (2)
% & $-$0.03 \fz (7) 
&    \fmb 0.01 \fz (1)
 & CLE
\\

%{{110, nlgp1}, 3, 8, 0, 100, 1000, 1000, 75454, 75454, 0.  0.25 I, 4, 1000, 1., 20., 10, 5.}Infinity
0.25i & $8^3$ % & 20 & 10
 & 19.749 \fz\fz (4) %&    0.001 \fz\fz (8)
% & \fm 0.000 \fz (2) 
&    55.969 \fz (5)
 & \fm 4.04 \fz (28) %&    0.04 \fz\fz (5)
% & $-$0.02 \fz (3) 
&  $-$  0.13 \fz (2)
 & CHB
\\
\hline

%{{53, rwmnlgp}}Infinity
0.5i & $3^3$ % &  & 
 & 71.642  (116) %&  $-$  6.146 \fz  (74)
% & \fm 8.487 (49) 
&    99.851  (47)
& \!\!\! $-$ 0.39 \fz (40) %&    0.53 \fz  (28)
% & \fm 0.08  (14) 
&  $-$  0.15  (21)
 & RW
\\

%{{2033, clnlgp}}Infinity
0.5i & $3^3$ % &  & 
 & 71.628 \fz  (76) %&  $-$  5.985  (127)
% & \fm 8.567 (49) 
&  \!\!\!\!  100.073 (98)
 & \fm 0.16 \fz  (75) %&  $-$  0.07 \fz  (15)
% & \fm 0.42 (22) 
&   \fmb 0.03 \fz  (9)
 & CLE
\\

%{{106, nlgp1}, 3, 3, 0, 100, 1000, 1000, 17464, 17464, 0.  0.5 I, 4, 1000, 1., 20., 10, 5.}Infinity
0.5i & $3^3$ % & 20 & 10
 & 71.578 \fz  (41) %&  $-$  6.173 \fz  (49)
% & \fm 8.510 (32) 
&    99.882 (32)
 & \fm 4.70 (125) %&    2.78 (135)
% & \fm 0.45 (22) 
&   \fmb 0.21 (29)
 & CHB
\\
\hline

%{{2037, clnlgp}}Infinity
0.5i & $8^3$ %&  & 
 & 71.332 \fz  (17) %&    0.076 \fz  (32)
% & $-$ 0.019   (18) 
&    99.579  (18)
 & \fm 0.24 \fz  (18) %&    0.06 \fz\fz  (4)
% & \fm 0.08 \fz  (6) 
&  $-$  0.01 \fz  (3)
 & CLE
\\

%{{1018, nlgp2}, 3, 8, 0, 100, 1000, 1000, 22, 22, 0.  0.5 I, 4, 1000, 1., 40., 12, 5., 0.}Infinity
0.5i & $8^3$ %& 40 & 12
 & 71.305 \fz\fz  (8) %&  $-$  0.014 \fz  (15)
% & \fm 0.004 \fz  (9) 
&    99.545 \fz  (9)
 & \fm 2.44 \fz  (85) %&    0.17 \fz  (19)
% & \fm 0.00 \fz  (5) 
&  $-$  0.01 \fz  (5)
 & CHB
\\

%{{1018, nlgp2}, 3, 8, 0, 100, 1000, 1000, 22, 22, 0.  0.5 I, 4, 1000, 1., 40., 12, 5., 0.}8
0.5i & $8^3$ %& 40 & 12
 & 71.305 \fz\fz  (8) %&  $-$  0.014 \fz  (15)
% & \fm 0.005 \fz  (9) 
&    99.544 \fz  (9)
 & \fm 0.14 \fz  (20) %&    0.02 \fz  (13)
% & \fm 0.00 \fz  (3) 
&   \fmb 0.03 \fz  (3)
 & CHB**
\\

%{{116, nlgp1}, 3, 8, 0, 100, 1000, 1000, 22753, 22753, 0.  0.5 I, 4, 1000, 1., 20., 10, 100000.}Infinity
0.5i & $8^3$ %& 20 & 10
 & 71.352 \fz  (11) %&    0.006 \fz  (13)
% & $-$0.008 \fz  (5) 
&    99.570 \fz  (8)
 & \fm 6.95 (212) %&  $-$  1.22 (264)
% & $-$0.06 (12) 
&  $-$  0.16 (13)
 & CHB*
\\

\hline

%{{2041, clnlgp}}Infinity
0.5i &$16^3$ %&  & 
 & 71.303 \fz\fz  (7) %&  $-$  0.014 \fz  (18)
% & $-$ 0.016 \fz  (7) 
&    99.569 \fz  (7)
 & \fm 0.11 \fz\fz  (7) %&    0.02 \fz\fz  (2)
% & $-$0.03 \fz  (3) 
&    \fmb 0.03 \fz  (1)
 & CLE
\\

%{{124, nlgp1}, 3, 16, 0, 100, 1000, 1000, 9416, 9416, 0.  0.5 I, 4, 1000, 1., 20., 10, 5.}Infinity
0.5i &$16^3$ %& 20 & 10
 & 71.328 \fz\fz  (5) %&  $-$  0.005 \fz\fz  (6)
% & \fm 0.000 \fz  (3) 
&    99.567 \fz  (4)
 & \fm 5.15 \fz\fz  (9) %&  $-$  0.09 \fz\fz  (7)
% & $-$0.01 \fz  (2) 
&  $-$  0.19 \fz  (3)
 & CHB
\\
\hline

%{{2043, clnlgp}}Infinity
0.5i & $8^4$ %&  & 
 & 90.652 \fz\fz  (9) %&  $-$  0.013 \fz  (12)
% & \fm 0.000 \fz  (4) 
&    94.365 \fz  (9)
 & \fm 0.06 \fz\fz  (9) %&  $-$  0.02 \fz\fz  (3)
% & \fm 0.00 \fz  (2) 
&    \fmb 0.03 \fz  (1)
 & CLE
\\

%{{126, nlgp2}, 4, 8, 0, 100, 1000, 1000, 65833, 65833, 0.  0.5 I, 4, 1000, 1., 20., 10, 5., 0.}Infinity
0.5i & $8^4$ %& 20 & 10
 & 90.670 \fz\fz  (5) %&    0.005 \fz\fz  (8)
% & \fm 0.000 \fz  (3) 
&    94.369 \fz  (5)
 & \fm 3.19 \fz  (20) %&  $-$  0.16 \fz  (16)
% & \fm 0.02 \fz  (2) 
&   \fmb 0.14 \fz  (3)
 & CHB
\\
\hline

\end{tabular}
\end{table}

As expected reweighting stops working for the larger lattices. Complex heat
bath works rather well for any lattice size for moderated couplings although
several technical issues arise, including the need of a cutoff and filtering
of rare (but catastrophic) cases (see \cite{Salcedo:2015jxd} for details).

Some small bias is observed in $\esp{I_1}$ (which ought to be zero). Likely,
this is rooted in a real weakness of the method, related to the presence of
\alert{zeroes in the marginal probabilities}. To analyze this, let us assume,
in a two-variable setting, that the current distribution $\rho$ is a proper
representation, $\esp{A}_\rho = \esp{A}_P$. One would like the Gibbs update to
preserve this property. A Gibbs update $(z_1,z_2) \to (z_1^\prime,z_2)$,
produces a new distribution $\rho^\prime$. Using the representation property
one has:
\begin{equation}\begin{split}
\esp{A}_{\rho^\prime}
&=
\int d^2z_2\, d^2 z_1^\prime \, \rho(z_2)  \,\rho_{\text{rep}}(z_1^\prime | z_2)
\, A(z_1^\prime , z_2)
=
\int d^2z_2\, d x_1^\prime \, \rho(z_2)  \frac{P(x_1^\prime, z_2)}{P(z_2)}
\, A(x_1^\prime , z_2)
\\
&=
\int d x_2\, d x_1^\prime \, P(x_2)  \frac{P(x_1^\prime, x_2)}{P(x_2)}
\, A(x_1^\prime , x_2)
+ \text{Residues}
= \esp{A}_P
+ \text{Residues}
.
\end{split}\end{equation}
So spurious contributions may arise from poles in $1/P(z_2)$. This idea is
fully sustained by a detailed analysis of
%\begin{equation}
$
P( x_1 , x_2 ) = (1 + \beta \cos(x_1) ) (1 + \beta \cos(x_2) )
(1 + \beta \cos(x_1 - x_2) )
.
$
%\end{equation}
A \alert{bias} is found whenever $[-Y,Y]$ contains the zero of $P(z_2)
\sim 1 + \frac{1}{2} \beta^2 \cos(z_2)$, and this cannot be prevented for
sufficiently large $\beta$ ~($Y$ increases and the zero $z_2$ decreases).

\begin{figure}[thb]\centering
\includegraphics[height=35mm,width=55mm]{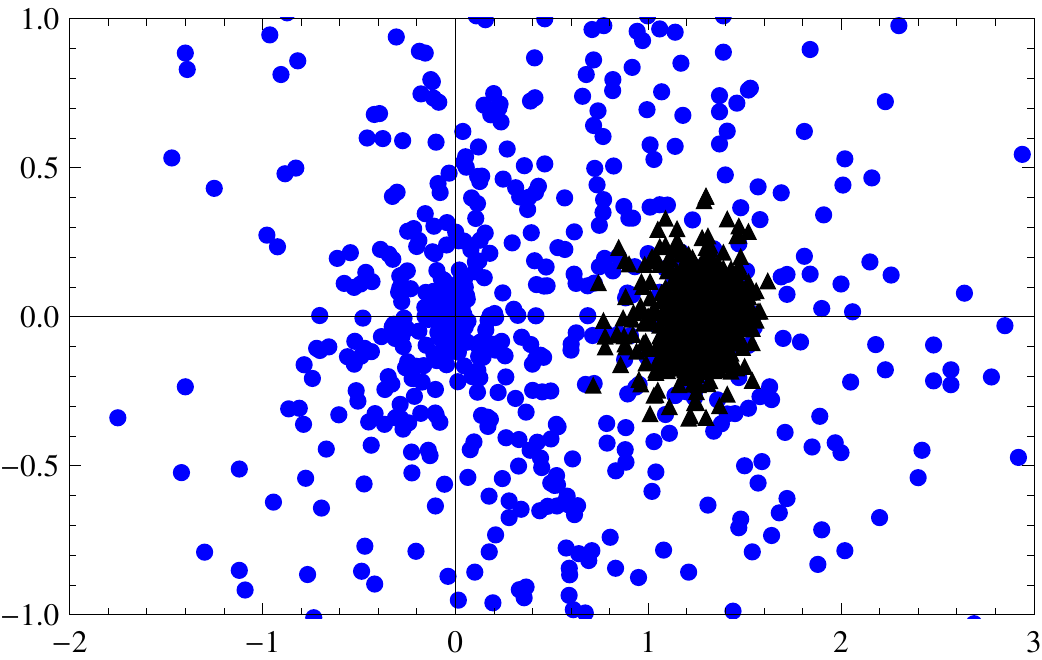}
\includegraphics[height=35mm,width=55mm]{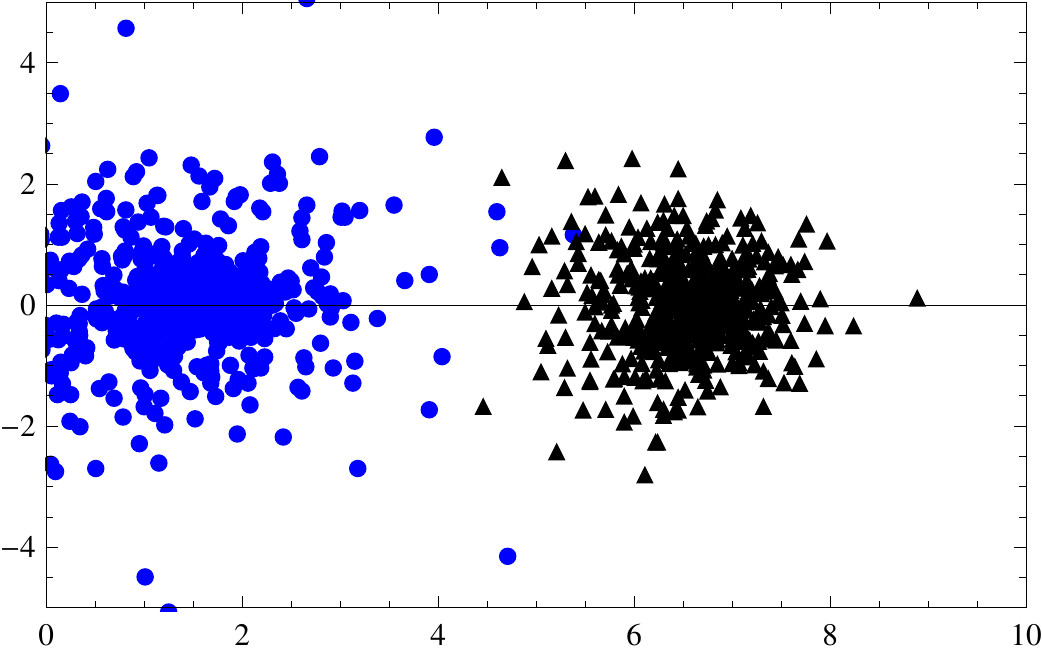}
\caption{Position of the marginal probability (or closely related quantity)
  obtained on-the-flight. Left: Model of Eq.~(\ref{eq:26}) for a cubic lattice
  and $\lambda=1$, for $\beta=0.5i$ (black triangles) and $\beta= i$ (blue
  disks).  Right: $XY$ model in an $8^3$ lattice, for $(\beta,\mu) = (1,1)$
  (black triangles), and $(\beta,\mu) = (0.4,0.2)$ (blue disks). As expected,
  one finds that only when the marginal probability stays away from zero
  (black triangles) the Gibbs sampling gives a fair estimate.}
\label{fig:chb-5}
\end{figure}
To have a robust method it is mandatory to find a way to remove the spurious
contributions from the poles in $1/P(z_1,\ldots,\widehat{z_i},\ldots,z_n)$,
when they exist. No such technique is available yet. On the other hand, it is
not hard to monitor on-the-flight whether the marginal probabilities stay away
from zero during the Markovian chain. This is displayed in
figure~\ref{fig:chb-5}.

\section{Summary and outlook}

Positive representations for any complex probability exist quite generally
with variable degrees of quality, related to their extension on the
complexified manifold. Good quality positive representations are expected to
exist for local actions, or more precisely, when the correlation length does
not increase with the size of the system. Unfortunately, there is no known
practical way to obtain them beyond the low dimensional case. CLE
would be such a construction when it works.

The complex heat bath method, which only needs one-variable representations,
works for mode\-rate couplings (moderately complex actions). The limitation is
due to the problem of possible zeroes in the marginal probabilities used to do
the updates. These zeroes introduce poles in the construction of the
representations of the conditional probabilities. The validity of the method
in each case can be assessed by monitoring the marginal probabilities
on-the-flight.  It would be worth studying whether the bias introduced by the
zeroes of the marginal probability could somehow be removed. In another
direction, perhaps the bias problem could be alleviated by considering larger
clusters of variables to be updated at each step (which is also more
costly).

Despite the impediments noted, may be the most interesting lesson to be
learned from the direct representation approach is that such positive
representations do exist, also in hard problems like QCD at finite density.
This means that there exist \alert{real} actions on the \alert{complexified}
manifold (however ugly, wide and non-local they could be) that correctly
reproduce such real-life systems. Then one can attempt a different route,
namely, to try to directly write down a real and \alert{local} action (on the
complexified manifold) in the same universality class as that of QCD at finite
density. Such real action would have the same symmetries (e.g., invariance
under \alert{real} gauge transformations) and would include a chemical
potential as an incorporated parameter but associated to the baryon number
conserved charge. In addition the action should be compatible with reflection
positivity (at least within the subset of holomorphic observables) and should
have a support with bounded width in the large volume limit. While this
does not seem to be an easy task, it should be recalled that, as we have
argued, such actions definitely exist, albeit without the locality and width
conditions.

%\clearpage
%\bibliography{lattice2017}

%%%%%%%%%%%%%%%%%%%%%%%%%%%%%%%%%%%%%%%%%%%%%%%%%%%%%%%%%%%%%%%%%%%%%%%%%%%%%
\end{document}